\documentclass{emulateapj}  
\usepackage[normalem]{ulem} 
\usepackage{verbatim}       

\slugcomment{Accepted for publication in the Astrophysical Journal, \textit{2012 Sep 30}}

\shorttitle{VISCOSITY AND WIND ON POLYTROPIC SELF-GRAVITATING DISKS}
\shortauthors{ABBASSI, NOURBAKHSH \& SHADMEHRI}

\begin{document}


\title{Viscous accretion of a polytropic self-gravitating disk \\ in the presence of wind}


\author{Shahram Abbassi\altaffilmark{1,2 $\star$}, Erfan Nourbakhsh\altaffilmark{1,2,3 $\dagger$},
and Mohsen Shadmehri\altaffilmark{4 $\ddagger$}}
\affil{$^1$ School of Physics, Damghan University, P.O. Box 36715-364, Damghan, Iran}
\affil{$^2$ School of Astronomy, Institute for Research in Fundamental Sciences (IPM), P.O. Box 19395-5531, Tehran, Iran}
\affil{$^3$ Department of Physics, Shahid Beheshti University, G.C., Evin, Tehran 19839, Iran}
\affil{$^4$ Department of Physics, Golestan University, Basij Square, Gorgan, Iran}


\email{$^\star$ E-mail: abbassi@ipm.ir (SA);}
\email{$^\dagger$ E-mail: e.nourbakhsh@mail.sbu.ac.ir (EN);}
\email{$^\ddagger$ E-mail: m.shadmehri@gu.ac.ir (MS);}


\begin{abstract}
Self-similar and semi-analytical solutions are found for the height-averaged equations
govern the dynamical behavior of a polytropic, self-gravitating disk under the effects of winds, around the nascent object.
In order to describe time evolution of the system, we adopt a radius dependent mass loss rate, then highlight its importance on both the traditional $\alpha$ and innovative $\beta$ models of viscosity prescription.
In agreement with some other studies, our solutions represent that Toomre parameter is less than one in most regions on the $\beta$-disk which indicates that in such disks gravitational instabilities can occur in various distances from the central accretor and so the $\beta$-disk model might provide a good explanation of how the planetary systems form.
The purpose of the present work is twofold.
First, examining the structure of disk with wind in comparison to no-wind solution;
and second, to see if the adopted viscosity prescription affects significantly the dynamical behavior of the disk-wind system.
We also considered the temperature distribution in our disk by a polytropic condition.
The solutions imply that, under our boundary conditions, the radial velocity is larger for $\alpha$-disks and increases as wind becomes stronger in both viscosity models. Also, we noticed that the disk thickness increases by amplifying the wind or adopting larger values for polytropic exponent $\gamma$. It also may globally decrease if one prescribe $\beta$-model for the viscosity. Moreover, in both viscosity models, surface density and mass accretion rate reduce as wind gets stronger or $\gamma$ increases.

\end{abstract}


\keywords{accretion, accretion disks --- stars: winds, outflows --- gravitation --- quasars: general}


\section{Introduction} \label{intro}

Accretion disk, a system where the disk feeds the central object through accretion, under the effect of viscous forces, is believed to be present in very different contexts and over a wide range of physical scales. In particular, such systems are found in active galactic nuclei (AGN), stellar black holes (X-ray binaries), young stellar objects (YSOs) and quasars (QSOs).
Since the angular momentum is strictly conserved, there needs to be a process that transports angular momentum away to prevent it from accumulating on the central object. The main acceptable possibility for this is the viscosity inside the disk (see Frank et al. 2002).
But the molecular viscosity is inadequate to transport angular momentum in the disk and describe luminous accretion disks, so some kind of turbulent viscosity is required.

Many investigators adopt the so-called $\alpha$-viscous model introduced by Shakura (1972) and Shakura $\&$ Sanyeav (1973) that gives the viscosity ($\nu$) at any radius ($r$) of the disk as a product of disk pressure scale height ($H$), the velocity of the sound ($c_{\rm s}$), and a parameter $\alpha$ that contains all unknown physics.
Despite a number of successful applications of the $\alpha$ prescription,
this parametrization suffers from a number of inconveniences (see Abbassi et al. 2006, hereafter AGS06).
Although the $\alpha$-prescription is based on a kind of turbulent viscosity
but there is no direct physical evidence for this as the origin of turbulence.


One of the most significant agents which may influence the equilibrium structure and dynamical evolution of almost any kind of disk, is \emph{self-gravity}. It is not merely a matter of forming systems. In the case of accreting super-massive black holes, for instance, during most of the evolution, they are self-gravitating.
Historically, accretion disk theory has concentrated on the non self-gravitating case, the effects of
self-gravity having only been discussed occasionally (e.g. Paczy\'{n}ski 1978; Kolykhalov \& Sunyaev 1979; Lin \& Pringle 1987, 1990).
For simplicity, traditional models of accretion disks assume geometrically thin configuration and neglect the self-gravity of the accreting material, which signifies that only pressure and the gravitational force of the central object support the vertical structure of the disk.
The study of disk self-gravity has been the subject of considerable attention in recent years.
It can be partly due to improved observations, which have shown that in several observed
systems the disk mass can be high enough to have a dynamical role on all scale disks, from AGN to protostars, and partly due to the increased computational resources, which have allowed a detailed numerical investigation of the development of gravitational instabilities in the non-linear regime (Lodato 2007 and references therein).

The structure of self-gravitating disks has been studied both through the self-similar solutions assuming steady and unsteady state (Mineshige \& Umemura 1996; Mineshige \& Umemura 1997, hereafter MU97; Mineshige et al. 1997, hereafter MNU97; Tsuribe 1999, hereafter TT99; Bertin \& Lodato 1999, 2001; Shadmehri \& Khajenabi 2006; AGS06; Shadmehri 2009, hereafter MS09), and through direct numerical simulations (Igumenshchev \& Abramowicz 1999; Stone et al. 1999; Torkelsson et al. 2000; Gammie 2001; Rice et al. 2003, 2005, 2010; Rice \& Armitage 2009; Cossins et al. 2010; Meru \& Bate 2011). Of these, a little attention has been paid to the polytropic flows (e.g. MNU97; AGS06).


Some laboratory experiments of Taylor-Couette systems (e.g., see, Richard \& Zahn 1999; Hur\'{e} et al. 2001) indicate that, although Coriolis force delays the onset of turbulence, the
flow is ultimately unstable to turbulence for \emph{Reynolds} numbers
larger than a few thousand. In all kind of self-gravitating disks the Reynolds number is extremely high, so it
was thought that probably the hydrodynamical driven turbulent
viscosity based on critical Reynolds number has a significant role
in the redistribution of angular momentum in the self-gravitating accretion disks.
The resulting hydrodynamically driven turbulence would then transport angular momentum
efficiently. Duschl et al. (2000) have proposed a generalized viscosity prescription
based on the hydrodynamically driven turbulence at the critical
effective Reynolds number, $\beta$-prescription, which can be applied for
both self-gravitating and non self-gravitating disks and is shown
to yield the standard $\alpha$-model in the case of non self-gravitating disks (Duschl et al. 2000).
They have demonstrated that in the case of fully self-gravitating
disks this model may reproduce very well the observed spectra of
proto-planetary disks and yield a natural explanation for the
accretion rate from the observed metallicity gradients in the disk
galaxy.
Furthermore, their study have presented that the standard model of thin accretion disk
based on $\alpha$ model leads to inconsistences if self-gravity plays an important role.
This problem arises from the parametrization of viscosity in terms of the local
sound speed and the vertical disk scale hight.

Following Duschl et al. (2000) suggestion for a $\beta$-prescription of viscosity,
AGS06 have applied this model for a thin axisymmetric, polytropic,
self-gravitating disk around a new born star.
Their results is quite different with standard $\alpha$ disks in
the outer part of the disks where the self-gravity becomes
important. But, in the inner part of the disks their solution converged to the standard $\alpha$ disks.


The next important possibility for angular momentum removal is an outflow or wind whose existence in many accreting systems is supported by strong observational evidences (e.g. Mobasher \& Raine 1989; Whelan et al. 2005; Bally 2007; Dionatos et al. 2009, 2010). Outflows are generally divided into two classes, called winds (poorly collimated) and jets (highly collimated)\footnote{In this work, we will mainly focus on winds, and the study of rather energetic winds or jets, whose speeds are sometimes comparable to that of light, is beyond the scope of the present study.}. The distinction between two classes is not always clear and in some cases, both are seen in the same object (see Murray 2002). It was long apparent that a disk wind/outflow contributes to loss of mass, angular momentum, and thermal energy from accretion disks (e.g. Piran 1977; Blandford \& Payne 1982; Pudritz 1985; K\"{o}nigl 1989).
It also appears to be an almost universal feature of disk-accreting systems on all astrophysical scales and frameworks.
The footprints of mass loss are observed around microquasars and YSOs as well as around massive stars (see chapter by Arce et al.) and even brown dwarfs (e.g. Ferrari 1998; Mirabel \& Rodr\'{\i}guez 1999; Bally et al. 2007; Whelan et al. 2005; Bourke et al. 2005), implying that the mechanism is of importance across the entire stellar mass spectrum (see review by Pudritz et al. 2007).
It is now widely accepted that winds or outflows have their origin in accretion flows (e.g. Blandford \& Payne 1982; Fender et al. 2004). Given that the majority of mass is accreted during earlier embedded phases, understanding disks at early times is more critical in a general sense. The initial winds could be blown from collapsar disk at these early phases, when the central YSO still has only a fraction of a solar mass (e.g., see, Banerjee \& Pudritz 2006, 2007).

The accretion flows lose their mass by winds as they accrete onto the central object. Here, it is found that the mass loss rate is dependent on radius of disk and stellar mass, and in order to study the dynamics of the disk-wind system, we shall describe the shape of this dependence by a power law. In the simplest standard model of star formation without wind (Shu 1977; Terebey et al. 1984), collapse proceeds in an \emph{inside-out} fashion, beginning in the center of the core, moving outward at the sound speed, and giving rise to a constant mass accretion rate of $\sim 2\times10^{-6}$ ${\rm M}_{\rm \odot}$ yr$^{-1}$. Of course, many modifications to this model have been also explored during last decades (Dunham \& Vorobyov 2012). As a result of mass loss, the accretion rate is no longer constant in radius, and has a power low dependance on it, with the power low index treated as a parameter of order unity (e.g. Blandford \& Begelman 1999; Abbassi et al. 2010). In the earliest phase of stellar evolution, average accretion rates are very high (nearly larger than $10^{-6}~{\rm M}_{\odot}~{\rm yr}^{-1}$) and disk winds accompany accretion with a wind mass loss rate that scales as $\sim \times 0.1$ the accretion rate (Gorti \& Hollenbach 2009).
Our understanding of the engine that powers the winds is, however, still limited.
As for the generation mechanisms of the disk winds, there are several important ingredients, such as electric field generated by the relative separation between ions and electrons, effects of magnetic fields, collisionless versus complications such as the effects of electron-positron pairs and coupling with the radiation field in the disk winds (Takahara et al. 1989).
Accordingly, to date, various driving sources of winds are proposed, including thermal, radiative, magnetic and other ones. Traditionally the name of wind depends on its driving force.


In vast majority of studies, authors readily suppose that the interstellar gas is isothermal and its pressure is proportional to the density. On account of isothermal collapse, the gas had to be assumed to already be clumpy and at very low temperatures (a few $\times$ 10 K). But undeniably, the first and most powerful break from this isothermality comes from protostellar radiation at the preliminary phases of star formation (Hansen et al. 2012). Examples include massive protostars which are capable of heating an entire cloud (Krumholz et al. 2007; Cunningham et al. 2011; Myers et al. 2011), and in lower level, low-mass protostars which of course may not have a same long-range influence as massive kinds.
In the present paper we presume the pressure to be proportional to the power of density (polytropic condition), instead of being proportional to the density alone (isothermal condition). Then, we compare the collapse of a polytropic gas disk with that of an isothermal one. We shall concentrate on the disk structure in the presence of rotating outflows or winds near newborn objects. However, since these winds are presumably launched by massive accretion disks (which is definitely true in early stages of protostellar evolution), we also pay heed to the possible influences of \emph{self-gravity} on disk-wind system.


Anyway, the influence of winds on accretion disks has been investigated by several authors (e.g. Meier 1979, 1982; Fukue 1989; Takahara et. al 1989; Knigge 1999, hereafter CK99; Xie \& Yuan, 2008; MS09; Kawabata \& Mineshige 2009; Dotan \& Shaviv
2011; Abbassi et al. 2010). In these studies, the accretion disk is usually height-integrated and
the configuration of the accretion flow is assumed rather than being calculated. Jiao \& Wu (2011) solved the full hydrodynamic equations to get the configuration of the accretion flow.
CK99 derived the radial distribution of dissipation rate and effective temperature across a Keplerian,
steady-state, mass losing accretion disk, using a simple parametric approach.
MS09 studied the influence of the winds on the time evolution of isothermal, self-gravitating accretion disks by adopting a power law mass loss rate because of the existence of wind. The work by MS09 considered both mass and angular momentum loss due to the wind/outflow, and provides a basis for the present work.

In accord with earlier discussion, viscosity, self-gravity and also outflows are among the most important physical agents in accretion-dominated stages of star formation. Although a considerable amount of work has gone into any pair of these, but no study has combined all three with non-isothermal circumstances using similarity approach. To our best knowledge, the simultaneous solutions for a viscous (with two prescriptions), self-gravitating, polytropic, mass-losing disk have not yet been reported.

Now, we are interested in considering the possibility that winds, could affect the global properties of polytropic self-gravitating accretion disks. Indeed, the present work is an attempt to provide a more thorough survey of solutions for two viscosity models with polytropic condition and wind, adopting different values of input parameters. As CK99 and MS09 the parametric model adopted here to describe the mass loss is simple, yet sufficiently general to be applicable to many types of dynamical disk-plus-wind models.
The paper organized as follows. In the next section we describe a basic physical approach for setting up analytical part of our model.
We did not consider the driving mechanisms of the wind.
In $\S$3 we firstly attempt to reduce our similarity equations in a slow accretion limit.
Then, assigning both proposed models for viscosity, we find two sets of ordinary differential equations and explore their solutions in $\S$4. A summery on the properties of our solutions and their implications are given in $\S$5.

\section{General Formulation}

\subsection{Basic equations}

Let us suppose a geometrically thin accretion disk
surrounding a central object which has not yet been completely formed. Thus, the radial component of the gravitational force is mainly provided by the self-gravity of the disk itself. The disk we are concerned with in this paper, has a symmetry over the rotation axis and is turbulent having an effective turbulent viscosity. Besides above-mentioned effects, we consider a wind emanating from disk surface, and set up fundamental governing equations in the cylindrical polar coordinates ($r,\varphi,z$) centered on accreting object with the equatorial plane of disk at $z=0$. The equations of continuity, radial momentum and conservation of angular momentum, respectively read (e.g. CK99; MS09)
\begin{equation}
\frac{\partial \sigma}{\partial t}
+\frac{1}{r}\frac{\partial}{\partial r}(r\sigma v_{\rm r})
+\frac{1}{2\pi r}\frac{\partial \dot{M}_{\rm w}}{\partial r}=0,
\label{eq:con}
\end{equation}
\begin{equation}
\frac{\partial v_{\rm r}}{\partial t}+v_{\rm r} \frac{\partial v_{\rm r}}{\partial r}
-\frac{v_{\rm \varphi}^2}{r}=-\frac{1}{\rho}\frac{\partial p}{\partial r}
+g_{\rm r}, \label{eq:eul}
\end{equation}
\begin{equation}
\sigma\frac{\partial (r v_{\rm \varphi})}{\partial t}
+\sigma v_{\rm r}\frac{\partial (rv_{\rm \varphi})}{\partial r}
=\frac{1}{r}\frac{\partial}{\partial r}\Big(r^3\sigma\nu\frac{\partial\Omega}{\partial r}\Big)
-\frac{(lr)^2 \Omega}{2\pi r}\frac{\partial \dot{M}_{\rm w}}{\partial r}, \label{eq:mom}
\end{equation}
 where $v_{\rm r}$, $v_{\rm \varphi}$, $\Omega$, $\rho$, $\sigma$, $p$ and $\nu$ are the radial, rotational and angular velocity, volume density, surface density, gas pressure and kinematic viscosity of the disk, in the order given. Other variables are set out in detail shortly afterwards.

Insofar as we require all quantities of the flow variables depend only on radius and time, we have integrated them in the vertical direction, following the thin disk approximation. So, rather than dealing with quantities per unit volume (such as the density $\rho$), we deal instead with quantities per unit surface (such as the surface density $\sigma=\int_{-\infty}^{+\infty} \rho dz$).

In Eq. (\ref{eq:eul}), we have radial component of the gravitational acceleration due to the self-gravity (cf. Saigo \& Hanawa 1998; TT1999),

\begin{eqnarray}
g_{\rm r} & = & -\frac{\partial \Phi}{\partial r} \nonumber \\
          & = & -2\pi G \int_{0}^{\infty}\int_{0}^{\infty} J_1(kr)J_0(kr')\sigma(r')kdkdr',
\label{eq:FG1}
\end{eqnarray}
derived from the axisymmetric Poisson equation adapted for a thin disk in cylindrical space (Nomura \& Mineshige 2000),

\begin{equation}
\\\frac{1}{r} \frac{\partial}{\partial r} \Big(r\frac{\partial \Phi}{\partial r} \Big)
+\frac{\partial^2 \Phi}{\partial z^2} = 4 \pi G \sigma(r) \delta (z).
\label{eq:Poisson}
\end{equation}
Here, functions $J_1$ and $J_0$ are the Bessel functions of the first kind and $\delta(z)$ is the Dirac delta function.
As usual, $G$ is the universal gravitational constant and $\Phi$ is the gravitational potential
contributed by the material of mass
\begin{equation}
\\M_{\rm r}(r)= 2\pi\int_0^r \sigma(r')~r' dr', \label{eq:Mr}
\end{equation}
inside of the disk within cylindrical radius $r$.

Since the gaseous disk is not rotating as a solid body, in the azimuthal equation of motion [Eq. (\ref{eq:mom})], we have the shear viscosity whose presence allows the transport of angular momentum from the faster inner
fluid elements to the slower outer ones.
Regardless of mechanisms which might contribute to the initiation of an outflow from the surface of an accretion disk, mass loss rate by wind/outflow is represented by $\dot{M}_{\rm w}$ in Eqs. (\ref{eq:con}) and (\ref{eq:mom}), and it is newly included in our analysis, whereas the \emph{Euler} Eq. (\ref{eq:eul}) does not change anymore despite inclusion of the wind. It is easy to perceive that merely the vertical component of wind velocity would make the gas escape from a surface of the disk. In fact, we have
\begin{equation}
\\\dot{M}_{\rm w}(r)=2 \int_{0}^{r} \dot{\sigma}_{\rm w}(r')~2\pi r' dr',
\label{eq:Mdotw}
\end{equation}
where $\dot{\sigma}_{\rm w}=\rho v_{\rm z}^{+}$ is mass loss rate per unit area from each disk face, picking $\rho=\rho|_{\rm z=0}$ as a midplane density of the disk and $v_{\rm z}^{+}(=-v_{\rm z}^{-})\geq0$ as mean vertical velocity at the disk surface, i.e. at the base of a wind. A factor of $\emph{2}$, appeared behind the integral, is for taking account of ejection from both disk faces.
Determining the wind velocity at its base is complicated because it depends on the vertical structure
of the disk which in turn depends on the detailed variation of the unknown viscosity with height. Furthermore,
the opening angle for the wind flow needs to be determined
using the geometry of the accreting flow and the pressure gradient in the wind in the radial direction (Misra \& Taam 2001).
Because $v_{\rm z}^{+}$ is unbeknown to us, we will compensate this by constructing a library of wind
solutions for a wide range of wind model parameters with $v_{\rm z}^{+}$ entangled therein.

The rightmost term of Eq. (\ref{eq:mom}) is the outflow sink added term represents angular momentum
transferred by the wind. Here, it is assumed that matter ejected at radius $r$ on the disk, carries away the angular momentum $(l r)^2\Omega$, where $\Omega = v_{\rm \varphi}/r$ is the angular frequency associated with lever-arm, $lr$, at radius $r$. Thus, $l =0$ corresponds to a non-rotating disk wind
and $l =1$ to outflowing material that carries away the specific angular momentum ($r^2\Omega$) it had right at the point of ejection.
This latter would be the most fitting value for radiation-driven outflows (e.g. Murray \& Chiang 1996; Proga et al. 1998; Feldmeier \& Shlosman 1999; Feldmeier et al. 1999).
Moreover, $0<l<1$ hints at the family of disk winds that carry away less angular momentum than possessed by the wind material before it left the disk surface.
Centrifugally-driven MHD disk winds (magnetocentrifugal winds for short) are corresponding to $l>1$ and can remove a lot of angular momentum from the disk (e.g. Blandford \& Payne 1982; Cannizzo \& Pudritz 1988; Emmering et al. 1992; Pelletier \& Pudritz 1992; Pudritz et al. 2007). In this case, we have $l=r_{\rm A}/r$, where $r_{\rm A}$ is Alfv\'{e}n radius (CK99). The angular momentum that is observed to be carried by rotating flows (e.g. DG Tau) is a consistent fraction of the excess disk angular momentum, from 60-100$\%$ (e.g. Bacciotti 2004), due to the high extraction efficiency  mentioned above. As we will discuss later in $\S$5, it is observationally well-known result that in many systems, $\dot{M}_{\rm w}/\dot{M}_{\rm acc} \sim 0.1$, with $\dot{M}_{\rm acc}$ being mass accretion rate which is introduced in $\S$\ref{sss}. This is faithful to the fact that lever-arm coefficients under the latter picture are often found in numerical and theoretical works to be $l=r_{\rm A}/r \sim 3$ - the observations of DG Tau being a perfect example (see a review by Pudritz et al. 2007).

To concede the truth of temperature distribution during the collapse, we employ a polytropic relation between pressure and density of accreting gas
\begin{equation}
\\p=K\rho^{1+\frac{1}{n}}=K\rho^{\gamma} \label{eq:poly}
\end{equation}
where $K$ is a constant set by the entropy of the gas and $n$ is known as the polytropic index for the process of interest. It is customary to replace $\gamma$ by $1+\frac{1}{n}$ and call it an effective adiabatic index. We should imply here that in actual accretion flows that can be represented by polytropes, neither $K$, nor $n$, may not be constant in space or time, and $\gamma (={d\ln p}/{d\ln\rho})$ does not necessarily equal $1+\frac{1}{n}$ (e.g. Goldreich \& Weber 1980; Yahil 1983). But we shall confine attention, in this paper, to an abstract case where each has certain fixed amount. The underlying polytropic exponent $\gamma$ is somehow an adjustment screw by which we can tune how much the temperature varies throughout the flow. The temperature increases more rapidly towards the center of polytropic gaseous disks with larger values of $\gamma$ (see MNU97). As $\gamma$ approaches unity, the temperature inclines to be uniform all around the disk, and consequently all values in the disk tend to those of isothermal case. The values in the disk such as geometrical thickness depends critically on $\gamma$ (see Saigo et al. 2000). According to Omukai \& Nishi (1998), the polytropic relation with $\gamma \approx 1.1$ is a good approximation to a collapsing metal-free gas cloud (see also Matsuda et al. 1969; Carlberg 1981).
If $c_{\rm s}$ is the local mean sound speed in barotropic disk, it satisfies
\begin{equation}
\\c^2_{\rm s}=\frac{dp}{d\rho}=K\gamma\rho^{\gamma-1}.
\label{eq:cs}
\end{equation}
Using last two equations, the pressure force per unit mass on the right-hand side of Eq. (\ref{eq:eul}) becomes
\begin{equation}
\\-\frac{1}{\rho}\frac{\partial p}{\partial r}=-\frac{c_{\rm s}^2}{\rho}\frac{\partial \rho}{\partial r}\approx-\frac{c_{\rm s}^2}{\sigma}\frac{\partial \sigma}{\partial r}, \label{eq:p2sgm}
\end{equation}
From the hydrostatic equilibrium equation in the vertical direction,
\begin{equation}
\\\frac{c_{\rm s}^2}{\sigma}\frac{\partial \sigma}{\partial z}+\frac{\partial \Phi}{\partial z}=0,
\end{equation}
and the Poisson Eq. (\ref{eq:Poisson}), we can get the vertical extent of the disk at any radius, as

\begin{equation}
\\H=\frac{c_{\rm s}}{(4\pi G\rho)^{\frac{1}{2}}}=\frac{c_{\rm s}^2}{2\pi G \sigma},
\end{equation}
viz. half-thickness of the disk, where we assume that the differential of the gravitational
potential in the radial direction is negligible compared with that in the vertical direction (e.g. Nomura \& Mineshige 2000). Here, as stated, the azimuthal integration leads to

\begin{equation}
\\\sigma(r)=\int_{-H}^{+H} \rho(r,z)~dz \approx 2\rho(r,0) H(r) .
\end{equation}

At last in this subsection, we apply the latest changes in our fundamental Eqs. (\ref{eq:con})-(\ref{eq:mom})
and rewrite them as
\begin{equation}
\frac{\partial \sigma}{\partial t}
+\frac{1}{r}\frac{\partial}{\partial r}(r\sigma v_{\rm r})
+2\dot{\sigma}_{\rm w}=0, \label{eq:con2}
\end{equation}
\begin{equation}
\frac{\partial v_{\rm r}}{\partial t}+v_{\rm r} \frac{\partial v_{\rm r}}{\partial r}
-\frac{j^2}{r^3}=-\frac{c_{\rm s}^2}{\sigma}\frac{\partial \sigma}{\partial r}
+g_{\rm r}, \label{eq:eul2}
\end{equation}
\begin{equation}
\frac{\partial j}{\partial t}
+v_{\rm r}\frac{\partial j}{\partial r}
=\frac{1}{r\sigma}\frac{\partial}{\partial r} \Big[r^3\sigma\nu\frac{\partial}{\partial r}(\frac{j}{r^2})\Big]
-\frac{2j}{\sigma}(l^2 \dot{\sigma}_{\rm w}),
\label{eq:mom2}
\end{equation}
where the angular momentum $j$ is replaced by $rv_{\rm \varphi}$, and we used Eq. (\ref{eq:Mdotw}) to obtain the radial derivative of mass loss rate, i.e. our auxiliary equation
\begin{equation}
\\\frac{\partial \dot{M}_{\rm w}}{\partial r}=4 \pi r \dot{\sigma}_{\rm w}. \label{eq:rrate}
\end{equation}

The solutions of recent four equations give us an inclusive overview of the evolution process
in infant stellar objects, which strongly depends on the chosen viscosity model,
and partly on the strength of existing wind.
So, the study of dynamical behavior of the accretion disks is postponed to
more information about the viscosity and also mechanisms through which the mass and angular momentum loses to the wind.

\subsection{Self-similar scaling} \label{sss}
The basic Eqs. (\ref{eq:con2})-(\ref{eq:mom2}) are troublesome to solve in real space, albeit approximately. For a sense of how they might be solved, it is useful to alter them to their non-dimensional forms, that firstly involves adopting an appropriate similarity variable. In this way, with the aid of $K$ and $G$ as the constitutive dimensional parameters, we pose a dimensionless similarity variable $x \equiv f_{\rm sc} r$, with
\begin{equation}
\\f_{\rm sc} = K^{-\frac{1}{2}}G^{\frac{\gamma-1}{2}}(\pm t)^{\gamma-2}
\end{equation}
 being the inverse length dimensional scale factor (e.g. MNU97, Yahil 1983, AGS06). Using such a combination of radius $r$ and time $t$, all hydrodynamic variables must therefore be functions of $x$ only and solutions will have the spatial structure at all times, because of self-similarity. Once the core formation epoch is an origin of time $(t=0)$, we take the plus sign in the parenthesis so as to gain positive time ($t>0$) solutions. Hence, bearing in mind the chain rule for the transformation $(r,t)\rightarrow (x,t^\prime)$, derivatives will turn into
\begin{equation}
\frac{\partial}{\partial r} \rightarrow K^{-\frac{1}{2}}G^{\frac{\gamma-1}{2}} (t^\prime)^{\gamma-2} \frac{\partial}{\partial x},
\end{equation}
\begin{equation}
\frac{\partial}{\partial t} \rightarrow \frac{\partial}{\partial
t^\prime}+(\gamma-2)\frac{x}{t^\prime}\frac{\partial}{\partial x}.
\end{equation}

From now on, we are allowed to write $t$ instead of $t^\prime$ and $d/dx$ instead of ${\partial}/{\partial x}$, since we require that all time-dependent terms should disappear in the self-similar forms of equations.
Now, it is convenient to draw up a list of useful real quantities correlated with their self-similar kinds:

\begin{equation}
\rho(r,t)=(4\pi\gamma)^{-\frac{1}{\gamma}}G^{-1}t^{-2}\Sigma^{\frac{2}{\gamma}}(x)
\end{equation}
\begin{equation}
p(r,t)=(4\pi\gamma)^{-1}KG^{-\gamma}t^{-2\gamma}\Sigma^2(x)
\end{equation}
\begin{equation}
c_{\rm s}(r,t)
=(4\pi)^{\frac{1-\gamma}{2\gamma}}\gamma^{\frac{1}{2\gamma}}K^{\frac{1}{2}}G^{\frac{1-\gamma}{2}}t^{1-\gamma}\Sigma^{\frac{\gamma-1}{\gamma}}(x)
\end{equation}
\begin{equation}
v_{\rm r}(r,t)=K^{\frac{1}{2}}G^{\frac{1-\gamma}{2}}t^{1-\gamma}V_{\rm r}(x)
\end{equation}
\begin{equation}
v_{\rm \varphi}(r,t)=K^{\frac{1}{2}}G^{\frac{1-\gamma}{2}}t^{1-\gamma}V_{\rm \varphi}(x)
\end{equation}
\begin{equation}
v_{\rm z}^{+}(r,t)=K^{\frac{1}{2}}G^{\frac{1-\gamma}{2}}t^{1-\gamma}V_{\rm z}^{+}(x)
\end{equation}
\begin{equation}
j(r,t)= KG^{1-\gamma}t^{3-2\gamma}J(x)
\end{equation}
\begin{equation}
\sigma(r,t)=(2\pi)^{-1}K^{\frac{1}{2}}G^{-\frac{1+\gamma}{2}}t^{-\gamma}\Sigma(x)
\end{equation}

\begin{equation}
H(r,t)=(4\pi)^{\frac{1-\gamma}{\gamma}}\gamma^{\frac{1}{\gamma}}K^{\frac{1}{2}}G^{\frac{1-\gamma}{2}}t^{2-\gamma}\Sigma^{\frac{\gamma-2}{\gamma}}(x)
\end{equation}
\begin{equation}
\nu(r,t)= KG^{1-\gamma}t^{3-2\gamma}\nu^\prime(x)
\end{equation}
\begin{equation}
M_{\rm r}(r,t)=K^{\frac{3}{2}}G^{\frac{1-3\gamma}{2}}t^{4-3\gamma}\mathcal{M_{\rm x}}(x)
\label{eq:Mrp}
\end{equation}

\begin{equation}
\dot M_{\rm r}(r,t)=K^{\frac{3}{2}}G^{\frac{1-3\gamma}{2}}t^{3(1-\gamma)} \dot {\mathcal{M}}_{\rm x}(x)
\end{equation}

\begin{equation}
\dot M_{\rm w}(r,t)=K^{\frac{3}{2}}G^{\frac{1-3\gamma}{2}}t^{3(1-\gamma)} \dot {\mathcal{M}}_{\rm w}(x)
\end{equation}

\begin{equation}
\dot{\sigma}_{\rm w}(r,t)=(4\pi)^{-1}K^{\frac{1}{2}} G^{-\frac{1+\gamma}{2}} t^{-\gamma-1}\Sigma(x)\Gamma(x)
\label{eq:mwp}
\end{equation}
where
\begin{equation}
J(x)=x V_\varphi(x),
\end{equation}
\begin{equation}
\mathcal{M_{\rm x}}(x)=\int_{0}^{x} \Sigma(x')x' dx', \label{eq:Mx}
\end{equation}
\begin{equation}
\dot {\mathcal{M}}_{\rm w}(x)=\int_{0}^{x} \Sigma(x')\Gamma(x')x' dx',
\end{equation}
\begin{equation}
\Gamma(x)=(4 \pi)^{\frac{\gamma-1}{\gamma}} \gamma^{-\frac{1}{\gamma}} \Sigma^{\frac{2(1-\gamma)}{\gamma}}(x) \Lambda(x)
~,~~\Lambda\equiv\Sigma V_{\rm z}^{+}.
\label{eq:GB}
\end{equation}
Also, the velocity of the constant $x$ surface seen from the rest frame $(r,t)$ is
\begin{displaymath}
v_{\rm rest}(r,t)= \frac{dr}{dt} \Big |_x = K^{\frac{1}{2}}G^{\frac{1-\gamma}{2}} t^{1-\gamma} V_{\rm rest}(x),
\end{displaymath}
\begin{equation}
V_{\rm rest}(x)= (2-\gamma)x.
\end{equation}
For later convenience, we define the comoving velocity as
\begin{equation}
V \equiv V_{\rm r}-V_{\rm rest}=V_{\rm r}+(\gamma-2) x. \label{V}
\end{equation}
It is useful to rewrite mass conservation Eq. (\ref{eq:con}) in terms of enclosed mass $M_{\rm r}$ [Eq. (\ref{eq:Mr})], i.e.
\begin{equation}
\frac{\partial M_{\rm r}}{\partial t}+ v_{\rm r}\frac{\partial M_{\rm r}}{\partial r}+\dot M_{\rm w}=0
~,~~\frac{\partial M_{\rm r}}{\partial r}=2\pi r \sigma.
\label{eq:cntyn}
\end{equation}
Then, introducing $\dot M_{\rm acc}$ as the mass accretion rate, we have $\dot M_{\rm acc}=\dot M_{\rm r}+\dot M_{\rm w}$, in such a way that
\begin{equation}
\\\dot M_{\rm acc}=-2\pi r \sigma v_{\rm r}.
\label{eq:Mdacc}
\end{equation}
A minus sign in this expression obviously shows when the radial velocity of the gas flow is directed inwards, an accretion may take place. Self-similarity leads to
\begin{equation}
\dot M_{\rm acc}(r,t)=K^{\frac{3}{2}}G^{\frac{1-3\gamma}{2}}t^{3(1-\gamma)} \dot \mathcal{M}_{\rm acc}(x),
\end{equation}
\begin{equation}
\dot \mathcal{M}_{\rm acc}=\frac{\Sigma x V_{\rm r}}{3\gamma-4}.
\end{equation}
%

\subsection{Basic equations in self-similar space}

After some replacements with the contribution of previous subsection, it is now possible to recast the differential Eqs. (\ref{eq:con2})-(\ref{eq:rrate}), from partial (PDEs) into ordinary ones (ODEs). The nondimensional similarity equations are then derived as

\begin{equation}
\frac{d}{dx} (\Sigma V x ) +(\Gamma-3\gamma+4)\Sigma x=0,
\label{eq:dcnty}
\end{equation}

\begin{equation}
V\frac{dV}{dx}
+\frac {2b^2}{\Sigma}\frac{d\Sigma}{dx}
-\frac{J^2}{x^3}
-\mathcal{G}_{\rm x}
-(2\gamma-3)V
+(\gamma-2)(\gamma-1)x=0,
\label{eqdrmo}
\end{equation}

\begin{equation}
V\frac {dJ}{dx}-\frac{1}{\Sigma x}\frac{d}{dx} \Big[\nu^\prime  x^3\Sigma \frac{d}{dx}\Big(\frac{J}{ x ^2}\Big) \Big]
+(l^2\Gamma-2\gamma+3)J=0,
\label{eq:dmom}
\end{equation}

\begin{equation}
\frac{d \dot\mathcal{M}_{\rm w}}{dx}= \Sigma x \Gamma,
\label{eq:drrate}
\end{equation}
where, for conciseness of notation, we introduced $b^2 = (4\pi\gamma)^{\frac{1-\gamma}{\gamma}} \Sigma^{\frac{2(\gamma-1)}{\gamma}}$, and the similarity gravitational field in radial direction is

\begin{equation}
\mathcal{G}_{\rm x}(x) = - \int_{0}^{\infty}\int_{0}^{\infty} J_1(kx)J_0(kx')\Sigma(x')kx'dkdx'.
\label{eq:FG2}
\end{equation}
Clearly, the effect of wind or outflows appears by the term $\Gamma$.
When we set this parameter equal to zero, the fourth equation can be left out and the other ones reduce to those appeared in MNU97 for $\alpha$-case or AGS06 for $\beta$-case (see $\S$\ref{subsec:vis}).
Although full numerical solutions to these equations would now be possible, it
is more instructive to proceed by analyzing the model in some restrictive cases such as one on the slow accretion limit.

\section{Reduction of Basic Equations}
\subsection{Helpful approximations}

In this subsection, we lay eyes on two useful approximations, and see how they reduce and simplify our equations, without much loss of generality of the problem.

Firstly, it seems favorable to ignore some terms like pressure gradient force and acceleration term in the cold and slow accretion limit which implies $V_\varphi \gg 1$, $\Sigma \gg 1$ and $|V_r|\ll 1$.
This estimate has been widely used by many authors to simplify formulas (e.g. MU97, MNU97, TT99, AGS06, MS09).
Furthermore, because in our model the wind velocity is expected to be smaller than radial inflow velocity,
one can handily assume $\dot {\mathcal{M}}_{\rm w} \ll 1$ as another implication of this limit, whose consistency with the results can be readily verified by reader. We will also check that $\dot M_{\rm w}$ do not exceed the accretion rate, $\dot M_{\rm acc}$. From a different but equivalent point of view, the slow accretion approximation is applicable for rotationally supported disks
when the viscous timescale is much longer than the dynamical timescale (TT99; AGS06).

Next, to avoid the integro-differential equation arises from substitution of Eq. (\ref{eq:FG2}) into  Eq. (\ref{eqdrmo}), we use a monopole approximation to compute the radial gravity field consistent with the mass distribution. Adopting this
approximation under the restrictions of the slow accretion limit, leaves (cf. TT99)
\begin{equation}
\\\mathcal{G}_{\rm x} \approx \frac{-\Sigma V}{(3\gamma-4)x}.
\label{eq:FG3}
\end{equation}
Except near the outer edge, this neglect of higher multipole orders is not expected to introduce any significant error as long as the surface density profile is steeper than $1/r$ (e.g. MU97; MNU97; Li \& Shu 1997; Saigo \& Hanawa 1998; TT99; Krasnopolsky \& K\"{o}nigl 2002; AGS06; MS09).

In the slow accretion limit, the third and fourth terms in Eq. (\ref{eqdrmo}) dominate and the others could be canceled (e.g. MNU97; AGS06). Then, by making use of last relation, we can demonstrate a radial force balance supplied by two terms in Eq. (\ref{eqdrmo}),
\begin{equation}
\\-\frac{J^2}{x^3}+\frac{\Sigma V}{(3\gamma-4)x}=0,
\label{eq:MJ}
\end{equation}
leading to
\begin{equation}
\\J= \Big(\frac{\Sigma V}{3\gamma-4}\Big)^{\frac{1}{2}} x.
\label{j1}
\end{equation}
One can take a logarithmic derivative of $J$ with respect to $x$, and get

\begin{equation}
\frac{d\ln J}{d\ln x}=
1+\frac {1}{2}\frac{d\ln \Sigma}{d\ln x}+\frac {1}{2}\frac{d\ln |V|}{d\ln x}.
\label{eq:dJ1}
\end{equation}
After some algebraic manipulations, Eq. (\ref{eq:dcnty}) takes the form

\begin{equation}
\frac{d\ln \Sigma}{d\ln x}=
-1-\frac{d\ln |V|}{d\ln x}+(3\gamma-\Gamma-4) \frac{x}{V},
\label{eq:dsigma}
\end{equation}
then can be substituted in Eq. (\ref{eq:dJ1}), to obtain

\begin{equation}
\frac{d\ln J}{d\ln x}=
\frac {1}{2}+\Big(\frac {3\gamma-\Gamma-4}{2V}\Big)x.
\label{eq:dJ}
\end{equation}

To proceed, from this point on, the kinematic coefficient of viscosity $\nu^{\prime}$
needs to be assigned as a function of our similarity variable $x$.

\subsection{Viscosity prescription} \label{subsec:vis}

For the viscosity $\nu'$ we shall apply two different
representations. The first and obvious one is the
$\alpha$-prescription introduced by Shakura \& Sunyaev (1973) and the resulting polytropic $\alpha$-disk plus wind will be formulated
in the next subsection. As mentioned in $\S$\ref{intro} the $\alpha$-viscosity is not a unique choice and
we also employ the so-called $\beta$-viscosity introduced by Duschl, Strittmatter \& Biermann (2000) thereinafter.

\subsubsection{$\alpha$-model solution}

In the case of $\alpha$-prescription, as suggested by MNU97 we adopt $\nu_{\alpha}^{\prime}=\alpha^{\prime}x^{\ell}$, with $\alpha^{\prime}$ and $\ell$ being free parameters. So, the viscosity coefficient is a function of $x$ only. Substituting this prescription into Eq. (\ref{eq:dmom}) we can inquire into a dynamical evolution of the disk.

Eqs. (\ref{eq:dsigma}) and (\ref{eq:dJ}) help us to simplify Eq. (\ref{eq:dmom}) and after some mathematical manipulations we can finally obtain a desired first order ODE for $\alpha$-viscous disks as

\begin{eqnarray}
\frac{dV_{\rm r}}{dx} \Big{|}_{\rm \alpha} & = & \frac{1}{\nu_{\rm \alpha}^\prime}\frac{[V_{\rm r}
+(2l^2-1)x\Gamma][V_{\rm r}+(\gamma-2) x]^2}{3V_{\rm r}-(3\gamma-2)x+2x\Gamma} \nonumber \\
&   & +\frac{Ax^2 + BV_{\rm r} x + 3(2\ell-1)V_{\rm r}^2 }{2x[3V_{\rm r}-(3\gamma-2)x+2x\Gamma]},
\label{eq:mainalpha}
\end{eqnarray}
where for a sake of brevity we have used
\begin{eqnarray*}
A & = & 4(2\ell+3)-4(\ell+3)\gamma-3\Gamma^2 \\
  &   & +[2\ell(\gamma-2)+6\gamma]\Gamma-2(2-\gamma)x\frac{d\Gamma}{dx}, \\
B & = & 6(\ell+2)\gamma-8(2\ell+1)+2(\ell-4)\Gamma+2x\frac{d \Gamma}{dx}.
\end{eqnarray*}

Clearly, the effect of wind/outflow appears by the term $\Gamma$. If we set this parameter
equal to zero, we can find the Eq. (34) of MNU97 which matches the solution without wind.
Considering $\ell=1$ and $\gamma=1$ for an isothermal case, the equation reduces to
Eq. (19) of MS09. If we continue by setting $\Gamma=0$, we could
easily obtain the Eq. (18) of MU97.

\subsubsection{$\beta$-model solution}

As we discussed in introduction we are also willing to use so-called $\beta$-prescription as
an alternative model for a self-gravitating disk which is introduced by Duschl et al. (2000) and is used by AGS06 in the form
\begin{equation}
\\\nu_{\beta}^{\prime}=\beta^\prime x V_{\varphi}=\beta^\prime J,
\end{equation}
with $\beta^{\prime}$ being a free parameter. It is worth noting that Duschl et al. (2000) $\beta$-viscosity recovers Shakura \& Sunyaev (1973) $\alpha$-viscosity for non-selfgravitating disks, if one requires the turbulence not to be super-sonic (which makes sense and is in agreement with Shakura and Sunyaev's parametrization, but still, it is an additional condition which one should mention). Likewise, with the aid of Eq. (\ref{eq:dJ}) and last equation, from Eq. (\ref{eq:dmom}) we can derive the single first order ODE for $\beta$-viscous disks as

\begin{eqnarray}
\frac{dV_{\rm r}}{dx} \Big{|}_{\rm \beta} & = & \frac{1}{\nu_{\rm \beta}^\prime}\frac{[V_{\rm r}
+(2l^2-1)x\Gamma][V_{\rm r}+(\gamma-2) x]^2}{3V_{\rm r}-(3\gamma-2)x+2x\Gamma} \nonumber \\
&   & +\frac{CV_{\rm r}+Dx}{3V_{\rm r}-(3\gamma-2)x+2x\Gamma}.
\label{eq:mainbeta}
\end{eqnarray}
where again for short we have written
\begin{eqnarray*}
C & = & 3(2-\gamma)+5(3\gamma-\Gamma-4)+x\frac{d\Gamma}{dx}, \\
D & = & (2-\gamma)(7\Gamma-x\frac{d\Gamma}{dx}-18\gamma+22)-2(3\gamma-4)^2.
\end{eqnarray*}
Same as previous, one may ignore the influence of wind in $\beta$-model, by adopting $\Gamma=0$, and this time recover the Eq. (32) of AGS06.

The ordinary differential Eqs. (\ref{eq:mainalpha}) and (\ref{eq:mainbeta}) along with Eq. (\ref{eq:drrate}) and Eq. (\ref{eq:dsigma}) are the main equations of our analysis which we can solve them numerically using the forth-fifth order Runge-Kutta-Fehlberg scheme. By exploit of asymptotic solutions as boundary conditions for our equations, we will investigate the effects of physical parameters on structure of the disk-wind system.

\section{Numerical Analysis}
In our model, there is a set of the input parameters like $s$, $l$, $\Lambda_0$, $\gamma$, ${\alpha}^\prime$ and ${\beta}^\prime$. Thus, a clear physical picture of our model is obtained only by an extensive parameter study. Here, we first obtain appropriate boundary conditions, and the possible effects of our different input parameters are explored in the subsequent subsections.

\subsection{Boundary conditions}
Our derived ODEs require boundary conditions so as to be solved numerically. Thus, at the first step, we derive asymptotic solutions for $V_r$. Our first limit is near the origin of the disk where $ x\rightarrow 0$, and the second one is in the outer part of the disk that, i.e. $x\rightarrow\infty$, which is toward the parent cloud. As a result for $\alpha$-model we have

\begin{eqnarray}
\lim_{x \rightarrow 0} V_{\rm r,\alpha} & \sim & -\frac{(4\ell+6)-(2\ell+6)\gamma}{3-2\ell} \nonumber \\
&   &\times \Big{[} x-\frac{9(4-3\gamma)x^3}{(2\ell-3)(6\ell^2-2\ell-15) \nu_{\rm \alpha}^\prime} \Big{]} \\
\lim_{x \rightarrow +\infty} V_{\rm r,\alpha} &\sim & -\frac{A \nu_{\rm \alpha}^\prime }{2x(\gamma-2)^{2}} - (2l^2-1)x\Gamma
\end{eqnarray}
Also for $\beta$-model it can be written as

\begin{eqnarray}
\lim_{x \rightarrow 0} V_{\rm r,\beta} & \sim & -\frac{5\gamma-6}{6\gamma-7}\Big{[}x -\frac{(\gamma-1)^2 x^3}{(6\gamma-7)\nu_{\rm \beta}^\prime}\Big{]} \\
\lim_{x \rightarrow +\infty} V_{\rm r,\beta} & \sim & - \frac{D {\nu_{\rm\beta}^\prime}}
{x(\gamma-2)^2} - (2l^2-1)x\Gamma
\end{eqnarray}
As $\gamma$ approaches unity, the results approach the isothermal case, again as expected.
Although to start the integration one can assume $V_{\rm r}=0$ at $x=0$, we should avoid the singularity by avoiding the origin. The above asymptotic behaviors helps us to determine an appropriate boundary conditions at a small neighborhood of singular point $x=0$. To have surface density profile, we can integrate surface density equation [Eq. (\ref{eq:dsigma})] from outer boundary, i.e. $x=1$, towards the center of the disk for a given $\Sigma(x=1)=\Sigma_{\rm out}$. Besides this, we demand a series of modes in which the accretion rate at the outer edge (i.e., the inflow rate from the parent cloud) is kept constant, which appears as a natural requirement (MS09). Thus, our main boundary conditions are $V_{\rm r}=0$ at $x=0$, and, $\dot{M}_{\rm acc}=\dot{M}_{\rm infall}$ at $x=1$. So, $\dot{M}_{\rm infall}$ is another input parameter to be assigned to our model. For the profile of mass loss rate, we prescribe a basic power law form for $\Lambda$ as $\Lambda =\Lambda_{0} x^{s}$, with $\Lambda_{0}$ and $s$ being the free parameters. Note that in the case of isothermal collapse, one derives $\Lambda=\Gamma$, and the wind parameters reduce to those of MS09.

\begin{figure*}
\vspace{0pt}
\hspace{13pt}
\includegraphics[width=16.43cm]{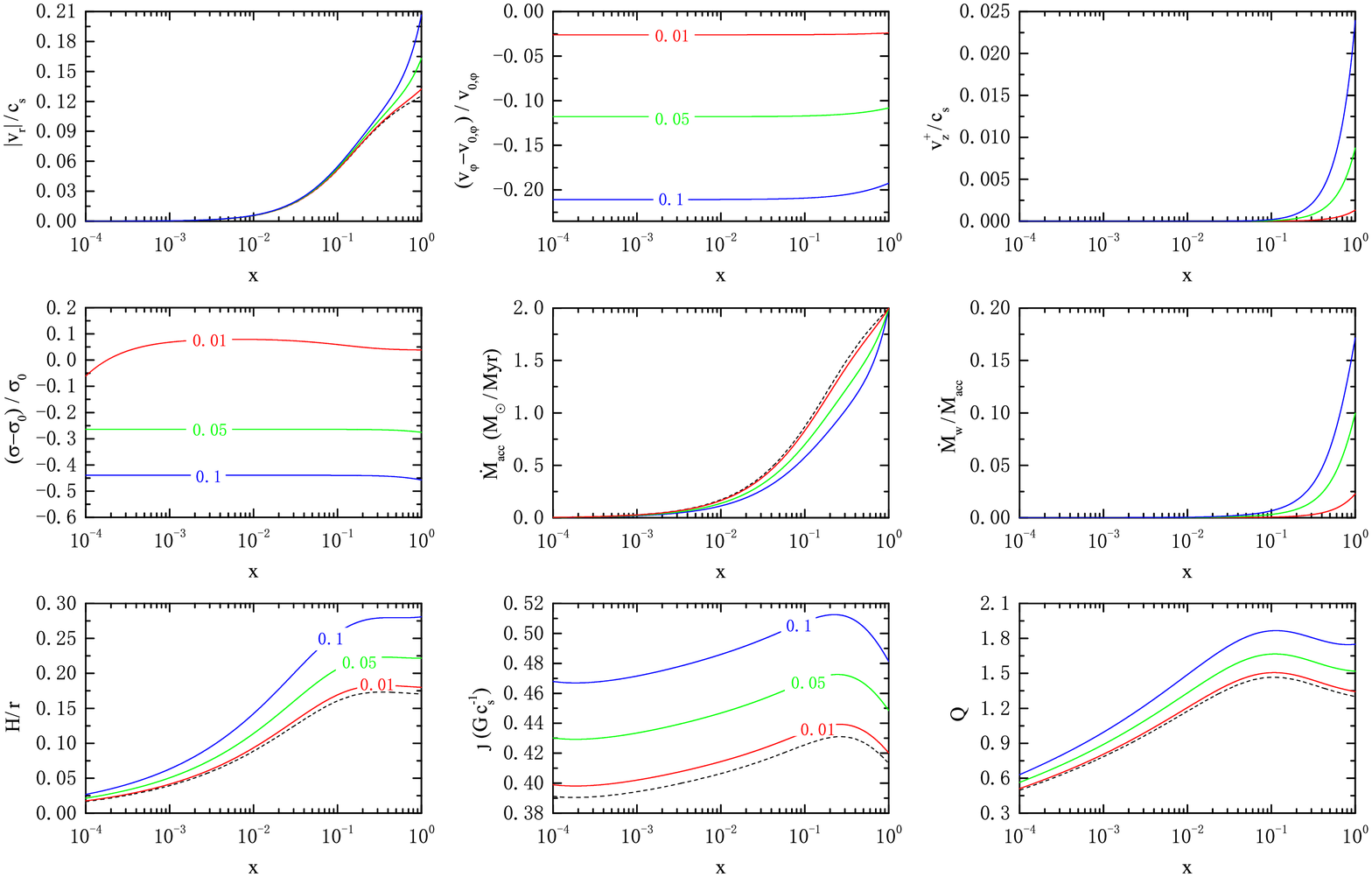}
\caption{The profiles of the physical variables for $\alpha' = 0.1$, $\ell = 1.0$, $\dot{M}_{\rm infall}=2.0 \times 10^{-6}~{\rm M}_{\odot}~{\rm yr}^{-1}$ and $\Lambda_{0}=0.1$, $0.05$, $0.01$  with $l=1.0$ and $s=0.7$ at $\gamma=1.1$. Surface density and the rotational velocity for no-wind solution are represented by $\sigma_{0}$ and $v_{0, \varphi}$. Each curve is labeled by corresponding $\Lambda_{0}$. No-wind solution is shown by dashed curves.}
\label{fig:alpha-Lambda0}
\end{figure*}

\begin{figure*}
\vspace{0pt}
\hspace{13pt}
\includegraphics[width=16.43cm]{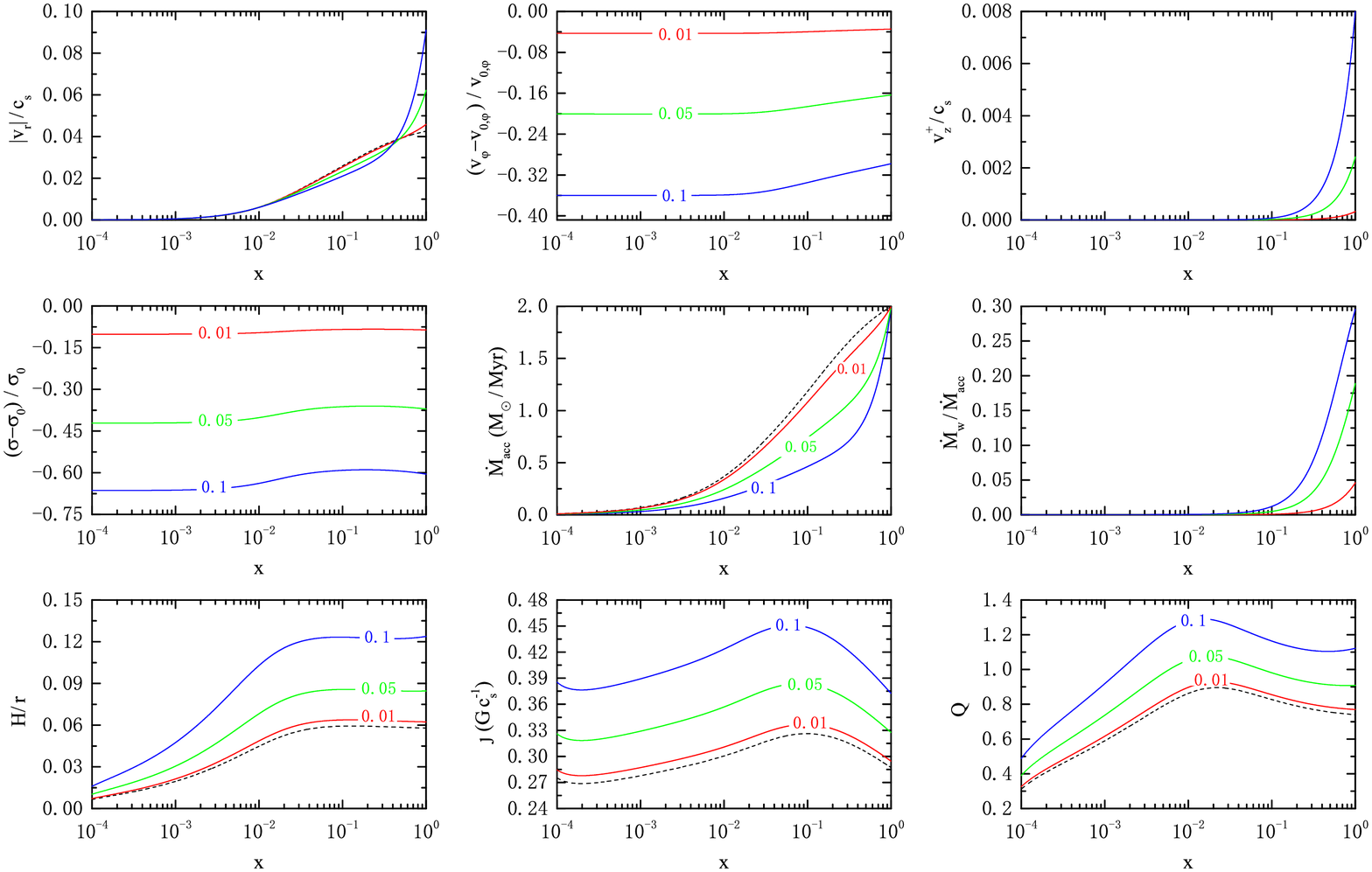}
\caption{The profiles of the physical variables for $\beta' = 10^{-3}$, $\dot{M}_{\rm infall}=2.0 \times 10^{-6}~{\rm M}_{\odot}~{\rm yr}^{-1}$ and $\Lambda_{0}=0.1$, $0.05$, $0.01$ with $l=1.0$ and $s=0.7$ at $\gamma=1.1$. Surface density and the rotational velocity for no-wind solution are represented by $\sigma_{0}$ and $v_{0, \varphi}$. Each curve is labeled by corresponding $\Lambda_{0}$. No-wind solution is shown by dashed curves.}
\label{fig:beta-Lambda0}
\end{figure*}

\begin{figure*}
\vspace{0pt}
\hspace{13pt}
\includegraphics[width=16.43cm]{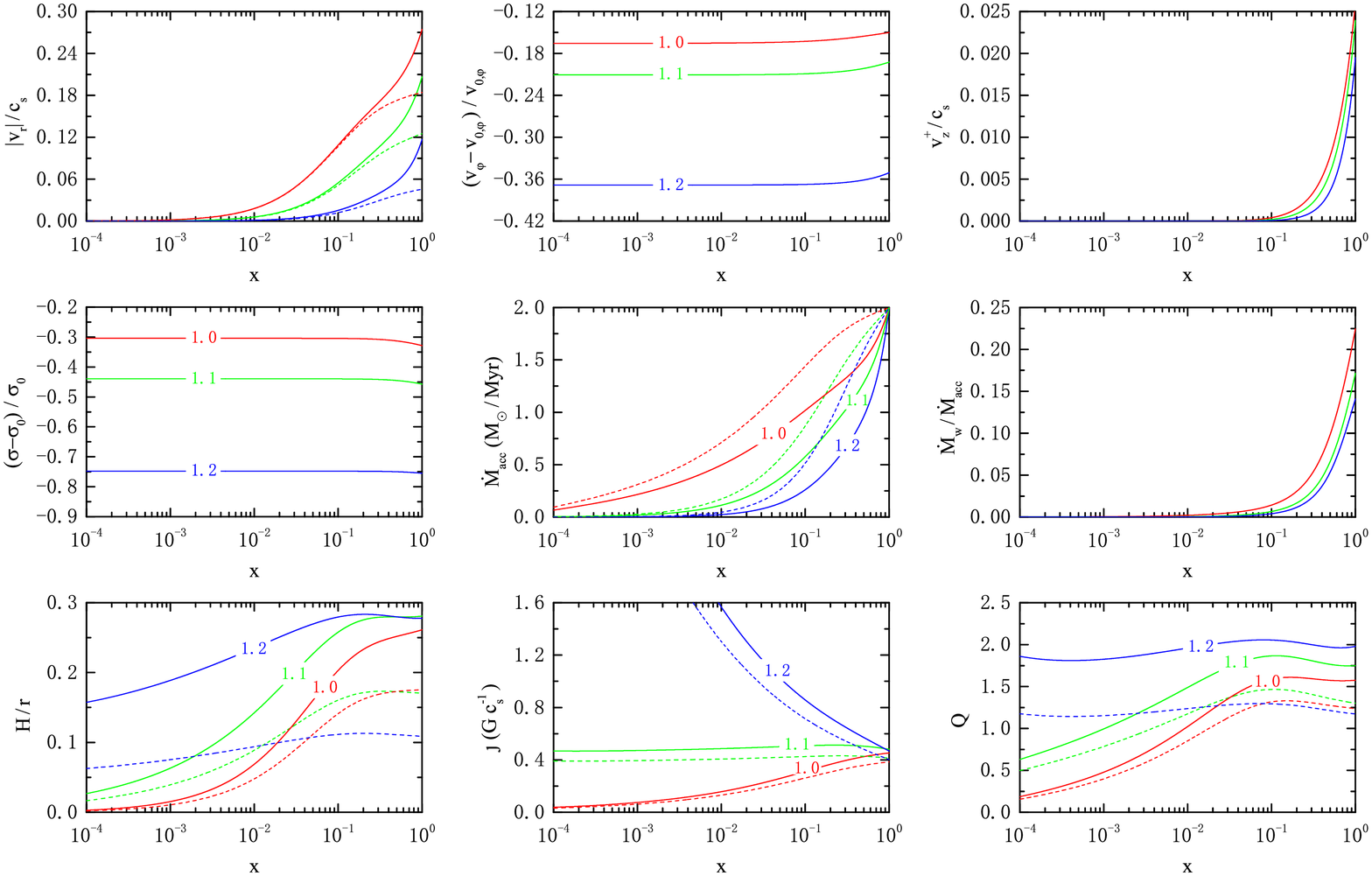}
\caption{The profiles of the physical variables for $\alpha' = 0.1$, $\ell = 1.0$, $\dot{M}_{\rm infall}=2.0 \times 10^{-6}~{\rm M}_{\odot}~{\rm yr}^{-1}$ and $\gamma=1.0$, $1.1$, $1.2$  with $\Lambda_{0}=0.1$, $l=1.0$ and $s=0.7$. Surface density and the rotational velocity for no-wind solution are represented by $\sigma_{0}$ and $v_{0, \varphi}$. Each curve is labeled by corresponding $\gamma$. No-wind solution is shown by dashed curves.}
\label{fig:alpha-gamma}
\end{figure*}

\begin{figure*}
\vspace{0pt}
\hspace{13pt}
\includegraphics[width=16.43cm]{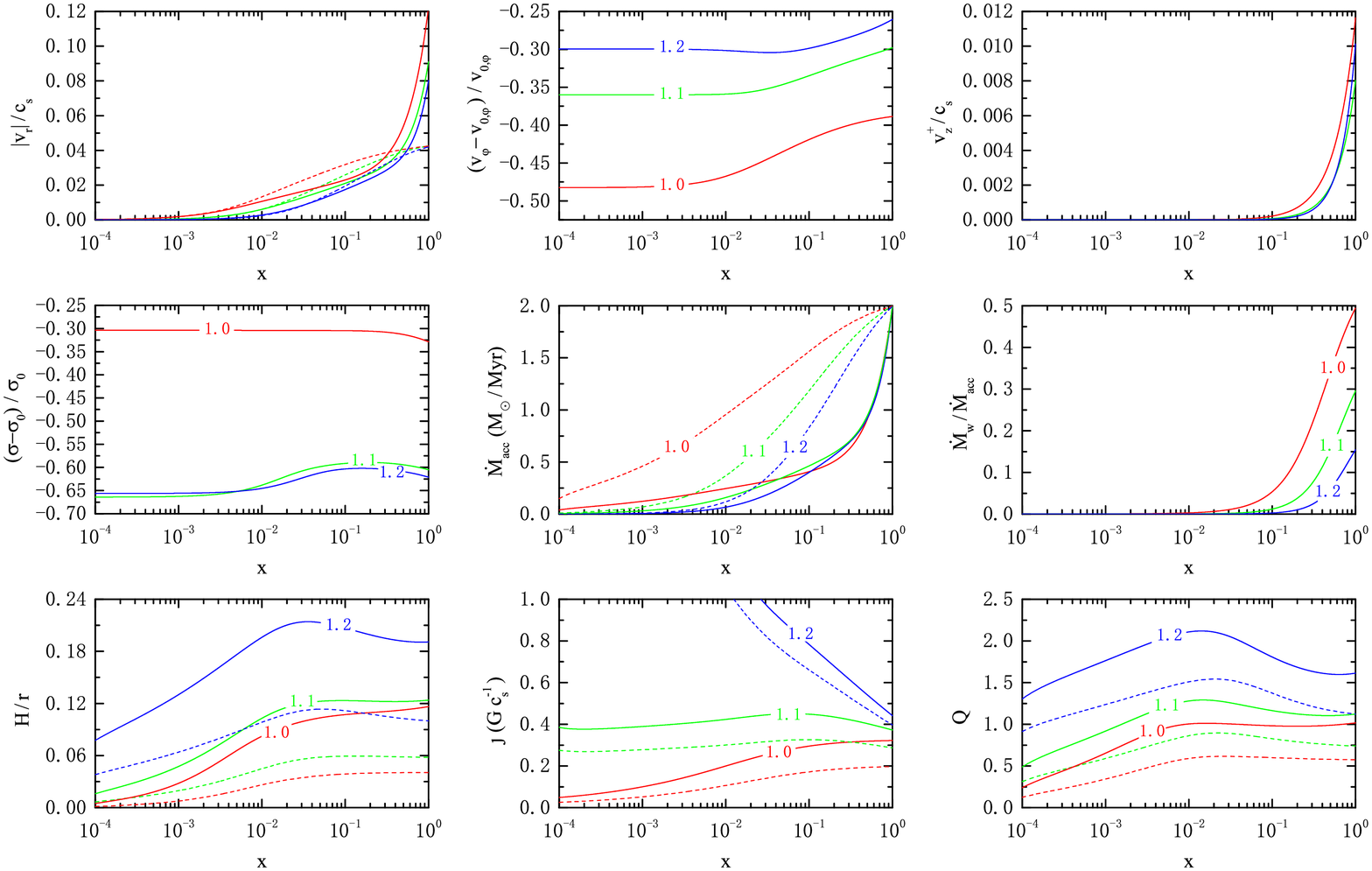}
\caption{The profiles of the physical variables for $\beta' = 10^{-3}$, $\dot{M}_{\rm infall}=2.0 \times 10^{-6}~{\rm M}_{\odot}~{\rm yr}^{-1}$ and $\gamma=1.0$, $1.1$, $1.2$ with $\Lambda_{0}=0.1$, $l=1.0$ and $s=0.7$. Surface density and the rotational velocity for no-wind solution are represented by $\sigma_{0}$ and $v_{0, \varphi}$. Each curve is labeled by corresponding $\gamma$. No-wind solution is shown by dashed curves.}
\label{fig:beta-gamma}
\end{figure*}

Now applying boundary conditions for our two systems of ODEs, we can obtain some profiles for hydrodynamic variables. Figs. \ref{fig:alpha-s}-\ref{fig:beta-gamma} show radial distributions for some important physical variables with parameterized values as a function of similarity variable $x$. For instance, one would immediately deduce that the radial velocity increases meaningfully at the outer part of the disk, because of the wind.
In order to make an easier comparison between surface densities, the ratio $(\sigma-\sigma_0)/\sigma_0$ versus
$x$ is shown in all figures and its negative value means that the surface density generally reduces in the presence of the disk wind.
In the figures, we see the mass accretion rate plainly decreases in comparison to the no-wind solution.
It is also an easy task to compare $\alpha$ with $\beta$ disks.
In the outer part of the disk - where the self-gravity is influential - the behavior of the solutions predicted by the $\beta$ viscosity model shows much less radial velocity compared to that by $\alpha$ viscosity model, either for wind or no-wind case. Moreover, it is apparent from the figures that winds could lessen the rotational velocity of the disk. Wind solutions imply that the amount of reduction to the rotational velocity is more significant for $\beta$-disks than it is for $\alpha$-disks, under our restrictive boundary conditions.

\subsection{Role of mass loss index $s$}
\begin{figure*}
\vspace{0pt}
\hspace{13pt}
\includegraphics[width=16.43cm]{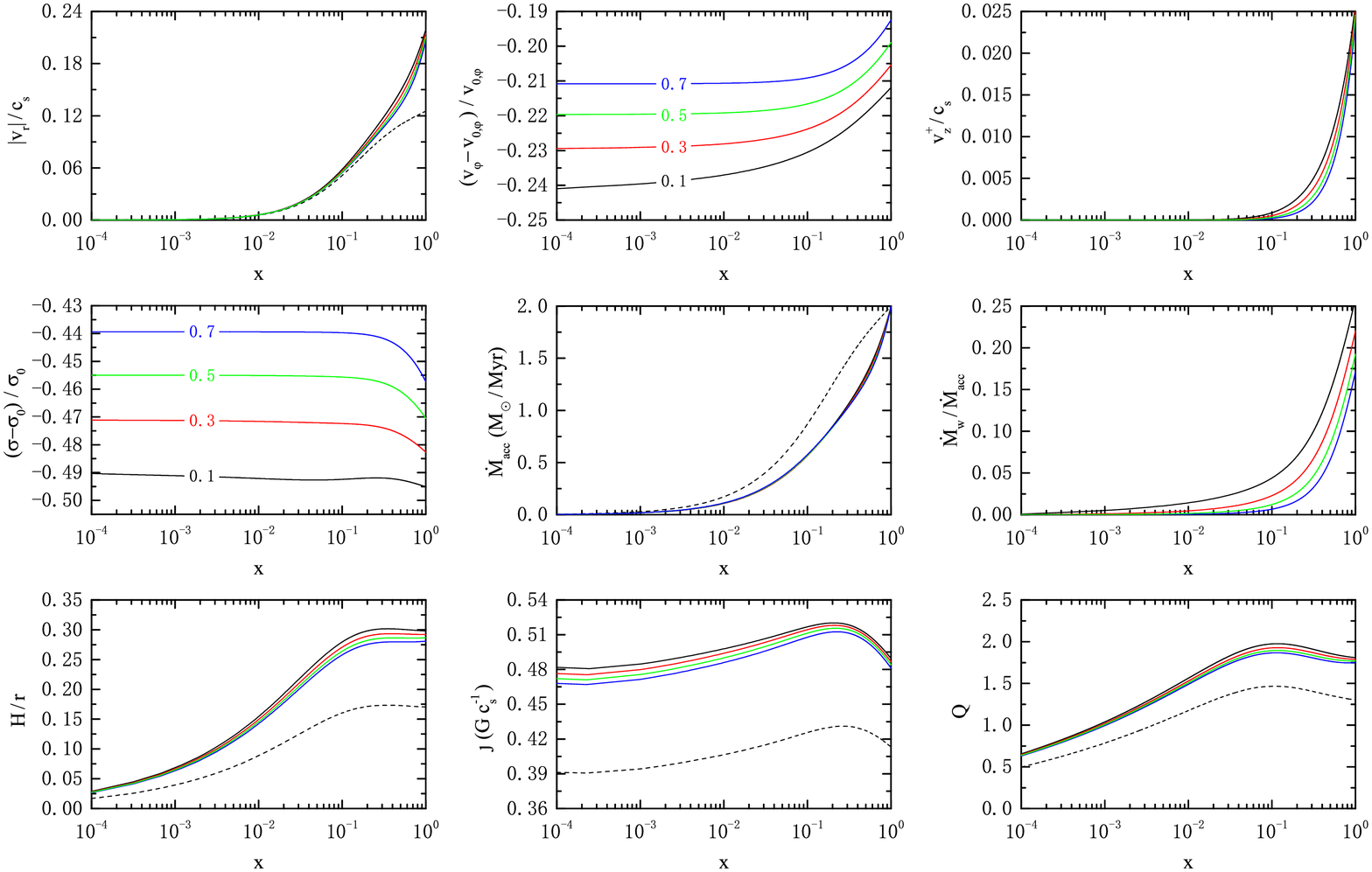}
\caption{The profiles of the physical variables for $\alpha' = 0.1$, $\ell = 1.0$, $\dot{M}_{\rm infall}=2.0 \times 10^{-6}~{\rm M}_{\odot}~{\rm yr}^{-1}$ and $s=0.1$, $0.3$, $0.5$, $0.7$  with $\Lambda_{0}=0.1$ and $l=1$ (i.e, rotating wind) at $\gamma=1.1$. Surface density and the rotational velocity for no-wind solution are represented by $\sigma_{0}$ and $v_{0, \varphi}$. Each curve is labeled by corresponding $s$. No-wind solution is shown by dashed curves.}
\label{fig:alpha-s}
\end{figure*}

\begin{figure*}
\vspace{0pt}
\hspace{13pt}
\includegraphics[width=16.43cm]{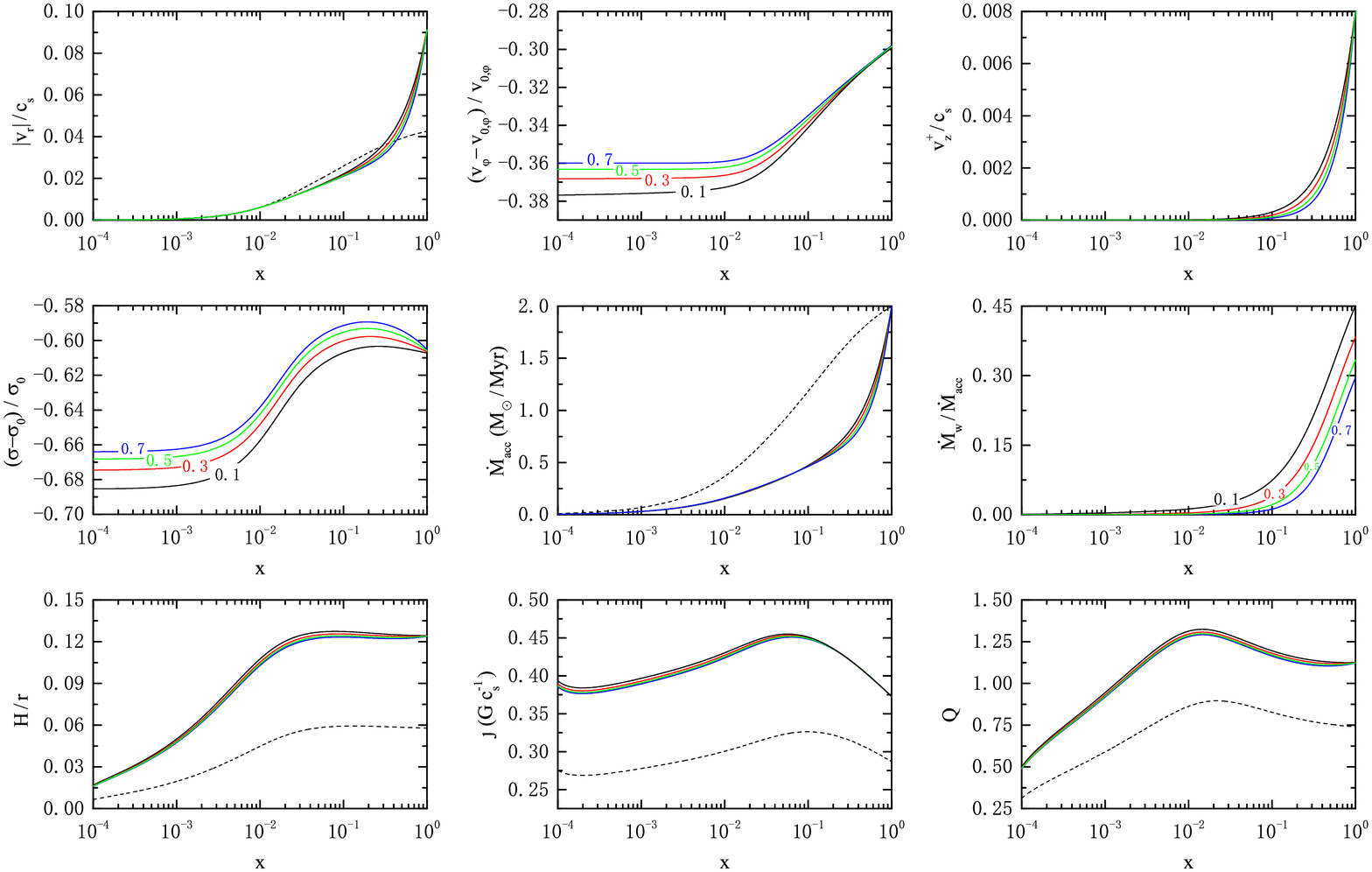}
\caption{The profiles of the physical variables for $\beta' = 10^{-3}$, $\dot{M}_{\rm infall}=2.0 \times 10^{-6}~{\rm M}_{\odot}~{\rm yr}^{-1}$ and $s=0.1$, $0.3$, $0.5$, $0.7$  with $\Lambda_{0}=0.1$ and $l=1$ (i.e, rotating wind) at $\gamma=1.1$. Surface density and the rotational velocity for no-wind solution are represented by $\sigma_{0}$ and $v_{0, \varphi}$. Each curve is labeled by corresponding $s$. No-wind solution is shown by dashed curves. }
\label{fig:beta-s}
\end{figure*}

Figs. \ref{fig:alpha-s} and \ref{fig:beta-s} show the effect of adopting various values for the mass loss power law index $s$, on the profiles of the physical variables. Each curve is labeled by corresponding index $s$. Also, we adopt $\alpha'=0.1$, $\ell=1$, $\beta'=10^{-3}$, $\dot{M}_{\rm infall}=2.0 \times 10^{-6}~{\rm M}_{\odot} / {\rm yr}$, $\Lambda_{0}=0.1$ and $l=1$ (i.e., rotating wind). Our adopted value for $\dot{M}_{\rm infall}$ is also compatible with the mean values inferred in embedded protostars (e.g. K\"{o}nigl \& Pudritz 2000). The surface density and the rotational velocity for the no-wind solution are represented by $\sigma_{0}$ and $v_{0,\varphi}$, respectively.
In order to make an easier comparison, the ratio $(\sigma - \sigma_{0})/\sigma_{0}$ as a function of variable $x$ is shown in Figs. \ref{fig:alpha-s} and \ref{fig:beta-s} (middle, left). Since the similarity radius is smaller than unity, for smaller values of parameter $s$ (consider $\Lambda=\Lambda_{0} x^{s}$), the wind becomes stronger and more mass is extracted from the disk. Therefore, surface density reduction is more significant for smaller values of $s$.

The rotational velocity profiles are presented as a function of the similarity variable in Figs. \ref{fig:alpha-s} and \ref{fig:beta-s} (top, middle). Generally, when the exponent $s$ decreases, the flow will rotate slower than that without winds. So, the viscous dissipation per unit mass in the flow is expected to be smaller in the presence of a wind. Also, the radial velocity profiles in Figs. \ref{fig:alpha-s} and \ref{fig:beta-s} (top, left) represent significant deviations from no-wind solution. In the outer parts of a disk the radial velocity is approximately uniform in the no-wind case. But, as the wind plays its crucial role at the outer parts of the disk, we have much larger radial velocity in comparison to the no-wind solution. Wind velocity $v_{z}^{+}$ at the surface of the disk is shown in Figs. \ref{fig:alpha-s} and \ref{fig:beta-s} (top, right). $\beta$-disks considerably have less $v_{z}^{+}$ than $\alpha$-disks. As wind gets stronger, its velocity at the surface of the disk increases which is quite expectable.

The accretion rate profiles, $\dot{M}_{\rm acc}$, are shown in Figs. \ref{fig:alpha-s} and \ref{fig:beta-s} (middle, middle). Accretion rate for the no-wind solution is represented by the dashed curves. Generally, the accretion rate decreases at all parts of the disk in the presence of the wind. Nevertheless, the accretion rate is not much sensitive to the variations of the exponent $s$. Ratio of the mass loss rate by wind to the accretion rate, i.e. the mass loss efficiency $\dot{M}_{\rm w}/\dot{M}_{\rm acc}$, is plotted in Figs. \ref{fig:alpha-s} and \ref{fig:beta-s} (middle, right). The mass loss due to the wind is negligible in the innermost region of the disk, except for strong winds, e.g. for $s=0.1$. Our input parameters were chosen so that mass loss efficiency is less than one at all radii of the disk. Here, the larger mass loss efficiencies appear at large radii, i.e. outer part of the disk. For a stronger winds which correspond to smaller values of $s$, a larger fraction of the mass carries away by the wind.

The profiles of disk aspect ratio, $H/r$, for various values of exponent $s$ are presented in Figs. \ref{fig:alpha-s} and \ref{fig:beta-s} (bottom, left). They demonstrate that in general $\alpha$-disks are thicker than $\beta$-disks and both of them are thicker in the presence of wind with all adopted values for $s$.
In Figs. \ref{fig:alpha-s} and \ref{fig:beta-s} (bottom, middle), we can see the distribution of angular momentum per unit mass, i.e. $j/M_{\rm r}=\jmath$, along the similarity radius of the disk, $x$, for different exponents $s$. As winds emanate from the disk, this fraction would be larger, particulary for smaller values of $s$. One would see the same behavior in a $\beta$-disk that typically has the lower values. Although there is an angular momentum loss due to the wind, the accompanying mass loss is high enough to keep $\jmath$ increasing as wind becomes stronger.

\subsection{Role of dimensionless lever-arm $l$}
\begin{figure*}
\vspace{0pt}
\hspace{13pt}
\includegraphics[width=16.43cm]{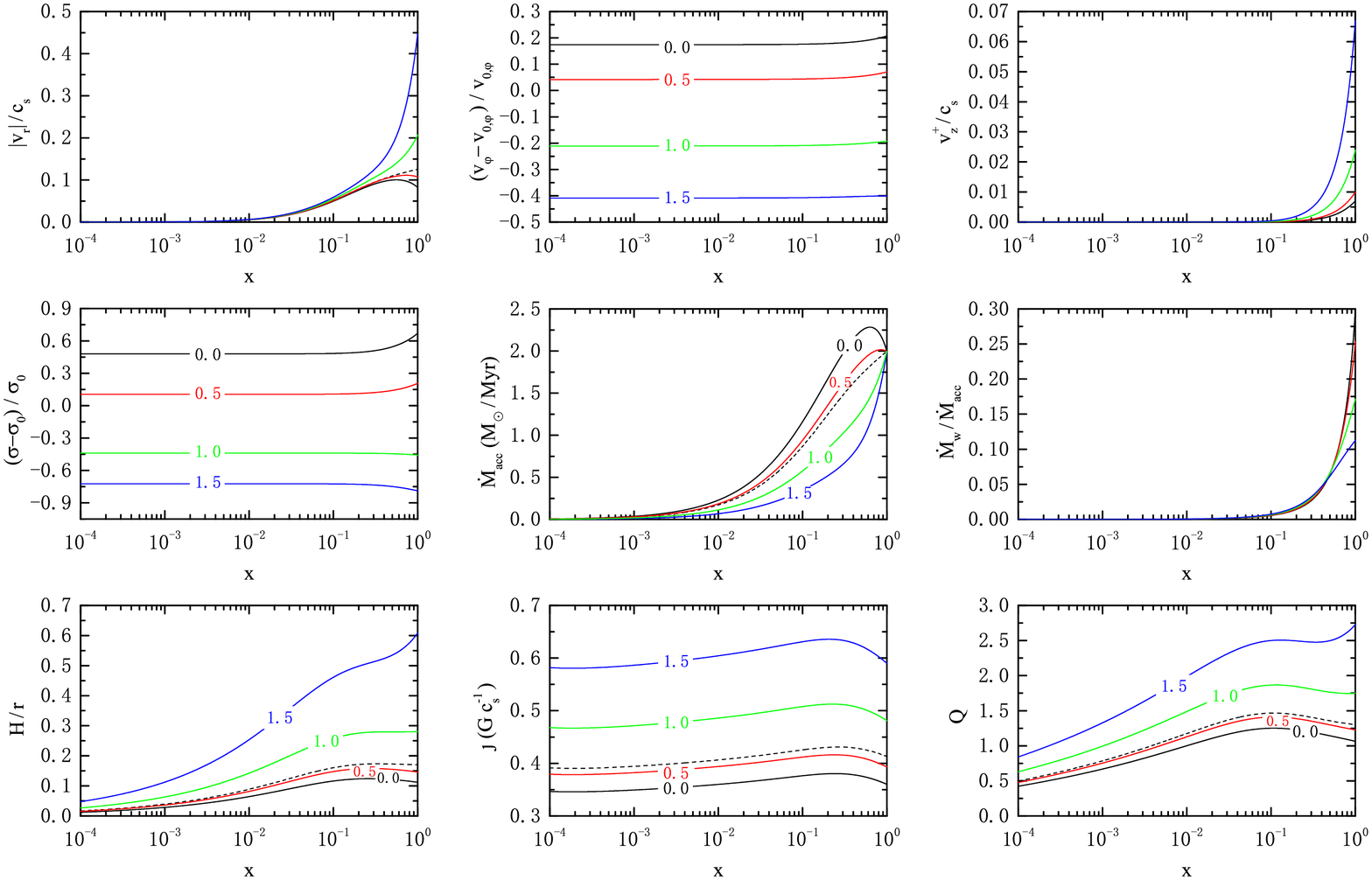}
\caption{The profiles of the physical variables for $\alpha' = 0.1$, $\ell = 1.0$, $\dot{M}_{\rm infall}=2.0 \times 10^{-6}~{\rm M}_{\odot}~{\rm yr}^{-1}$ and $l=0.0$, $0.5$, $1.0$, $1.5$  with $\Lambda_{0}=0.1$ and $s=0.7$ at $\gamma=1.1$. Surface density and the rotational velocity for no-wind solution are represented by $\sigma_{0}$ and $v_{0, \varphi}$. Each curve is labeled by corresponding $l$. No-wind solution is shown by dashed curves.}
\label{fig:alpha-l}
\end{figure*}

\begin{figure*}
\vspace{0pt}
\hspace{13pt}
\includegraphics[width=16.43cm]{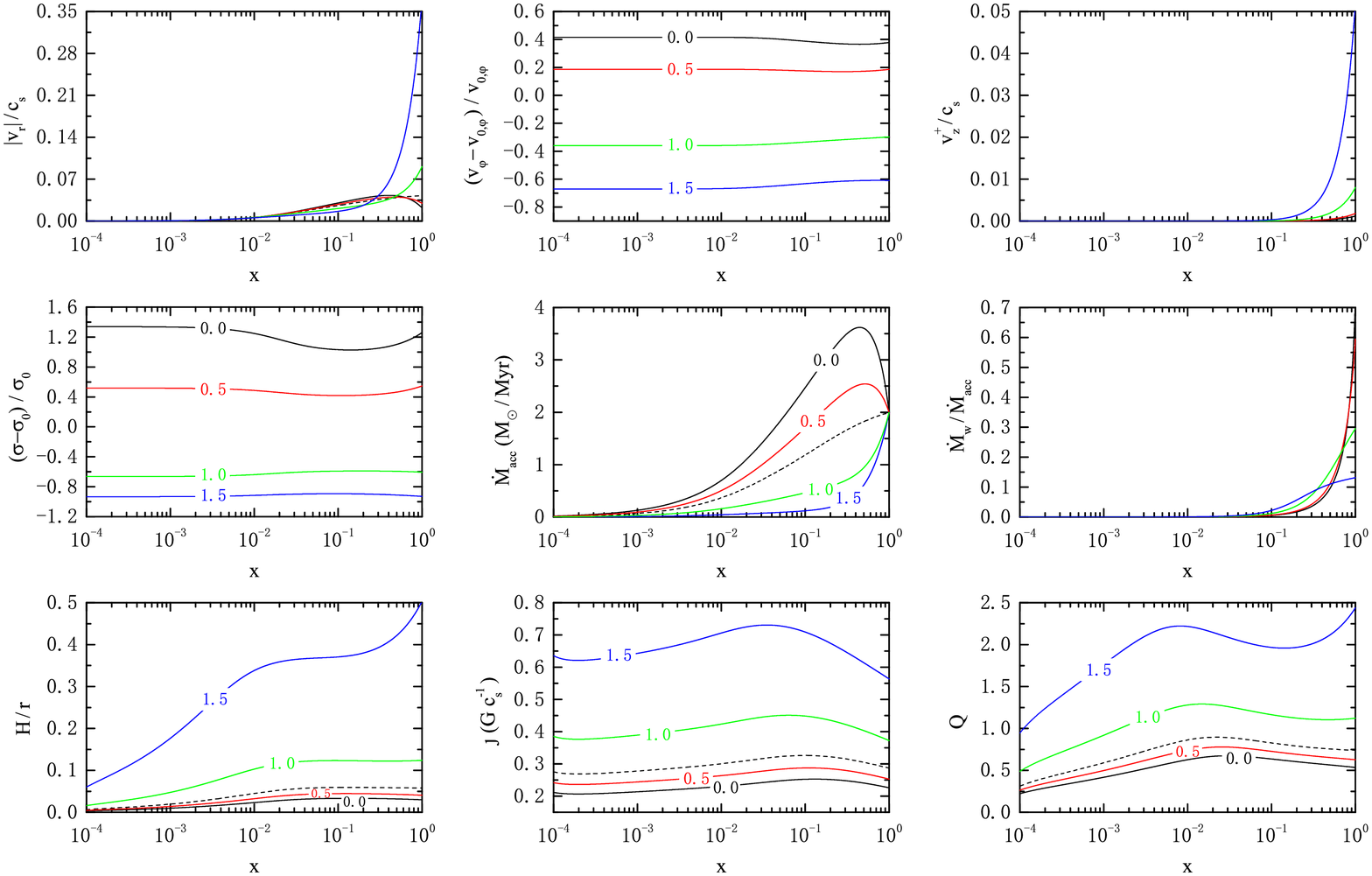}
\caption{The profiles of the physical variables for $\beta' = 10^{-3}$, $\dot{M}_{\rm infall}=2.0 \times 10^{-6}~{\rm M}_{\odot}~{\rm yr}^{-1}$ and $l=0.0$, $0.5$, $1.0$, $1.5$  with $\Lambda_{0}=0.1$ and $s=0.7$ at $\gamma=1.1$. Surface density and the rotational velocity for no-wind solution are represented by $\sigma_{0}$ and $v_{0, \varphi}$. Each curve is labeled by corresponding $l$. No-wind solution is shown by dashed curves.}
\label{fig:beta-l}
\end{figure*}

Possible effects of an angular momentum extraction due to the wind are explored in Figs. \ref{fig:alpha-l} and \ref{fig:beta-l} by adopting various values of the input parameter $l$. We here assume $\alpha' = 0.1$, $\ell = 1.0$, $\beta'=10^{-3}$, $\dot{M}_{\rm infall}=2.0 \times 10^{-6} M_{\odot} / {\rm yr}$ and $l=0.0$, $1.0$, $1.5$  with $\Lambda_{0}=0.1$ and $s=0.7$. Obviously, when we have $l=0$, angular momentum is not extracted by the wind. This case corresponds to a non-rotating wind and the disk losses only mass because of the wind. However, as mentioned in MS09, it can be shown that for $l^{2}<1/2$ the mass of the disk increases in the presence of the winds that obviously has not any physical meaning. This is partly due to the limitations of similarity method that there is not a self-consistent solution for any given set of the input parameters. More importantly, our model is valid just in the slow accretion limit which implies $V_{\varphi} \gg 1$, and so it is very unlikely to accept that winds are lunched without extracting a certain amount of angular momentum of the disk (MS09). Although we have represented solutions with $l=0,0.5$ in Figs. \ref{fig:alpha-l} and \ref{fig:beta-l} for a sake of comparison, as in MS09, we think these solutions are not physically acceptable. Profiles of surface densities for each viscosity model are shown in Figs. \ref{fig:alpha-l} and \ref{fig:beta-l} (middle, left). We can see again the reduction to the surface density because of the wind. Rotational velocity of the disk decreases because of the angular momentum removal from the disk, as $l$ becomes larger (top, middle in figures). Although, the radial velocity in the innermost part the disk does not change because of the wind, but in comparison to the no-wind solution, existence of a rotating wind enhances the radial velocity at the outer part of the disk (top, left). The typical behavior of wind velocity $v_{\rm z}^{+}$ is also sensitive to the amount of the extracted angular momentum (top, right). The accretion rate profile (middle, middle) represents that it decreases due to the existence of a rotating wind. However, for a large $\l$, where more angular momentum is carried away by the wind, as long as the surface density and the rotational velocity are reduced at all regions of the disk, the radial velocity of the accretion flow at the outer part of the disk is increased significantly (cf. MS09). We also plotted the specific disk angular momentum $\jmath$ as a function of $x$ so that one can see its behavior for $\alpha$ and $\beta$ disks assuming $\gamma=1.1$ for different values of $\l$. Here, it is informative to compare solutions for $l=0,0.5$ with those for $l=1.0,1.5$. It can be inferred from the disk aspect ratio profile (bottom, left) that the more angular momentum is removed from the disk, the thicker it gets.

\subsection{Role of factor $\Lambda_0$}
One of the prominent input parameters in our model is $\Lambda_0$ that its possible effects are explored in Figs. \ref{fig:alpha-Lambda0} and \ref{fig:beta-Lambda0}. We assume that $\alpha' = 0.1$, $\ell = 1.0$, $\beta'=10^{-3}$, $\dot{M}_{\rm infall}=2.0 \times 10^{-6} M_{\odot} / {\rm yr}$ and $\Lambda_{0}=0.1$, $0.05$, $0.01$  with $l=1.0$ and $s=0.7$. The surface density and the rotational and radial velocities are substantially decreasing with $\Lambda_0$. The wind velocity at the surface of the disk is significantly affected by the parameter $\Lambda_0$ (top, right). As a result, the mass accretion rate and the wind mass loss rate are respectively decreased and increased with the parameter $\Lambda_0$. Further, $\jmath$ increases in both viscosity models, as wind becomes stronger by adopting larger values of $\Lambda_0$.

\subsection{Effect of disk self-gravity} \label{subsec:q}

The effect of the disk self-gravity in this paper is limited to provide the radial gravitational field to keep the disk in centrifugal equilibrium. On the other hand, it is predicted that in the outer part of accretion disks around QSOs, self-gravity has a dominant role. This effect is investigated by Toomre (1964). As a simplest indicator for gravitational stability of the solutions we can use the Toomre criterion,
\begin{equation}
Q=\frac{c_{\rm s} \kappa}{\pi G\sigma}
\end{equation}
where
\begin{equation}
\kappa=\Omega(4+2\frac{d\log\Omega}{d\log r})^\frac{1}{2}
\end{equation}
is the epicyclic frequency at which a fluid element oscillates when perturbed from circular motion.
In a nearly Keplerian disk, $\kappa \approx \Omega$. For axisymmetric disturbances,
disks are stable against the gravitational fragmentation when $Q > 1$. The local gravitational instability occurs when $Q < 1$.
Now, we rewrite our gravitational instability parameter in the self-similar form as
\begin{equation}
Q=2\sqrt{2} (4\pi)^{\frac{1-\gamma}{2\gamma}}
\gamma^{\frac{1}{2\gamma}} \Sigma^{\frac{-1}{\gamma}} x^{-2} J \Big (\frac{d\ln J}{d\ln x} \Big )^{\frac{1}{2}},
\end{equation}
which by setting $\gamma=1$, we recover Eq. (18) of TT99, viz. $Q=(2\sqrt{2} J \sqrt{d\ln J / d\ln x} )/ \Sigma  x ^2$.
In all figures (bottom, right), we have shown the distribution of the Toomre Q-value for
some parameters.

To make an easier comparison, Toomre parameter for a case without wind/outflow is also represented. The solutions indicate that Toomre parameter increases with winds or outflows, except for the cases with $l=0, 0.5$ which gives unphysical solutions, as we discussed previously. Generally in $\alpha$-model, except for the inner part of the disk, Toomre parameter is still larger than one, especially when the winds are present. However, in the case of $\beta$-prescription for viscosity, we see that $Q$ is below the instability threshold (i.e. $Q_{\rm thr}\approx1$) in most regions of the disk, even in the presence of a typical wind (but some cases, e.g. for $l=1.5$). Manifestly, one will not have any trouble adjusting the input parameters for $\beta$-disk to be locally unstable in various distances from the central accretor and so the $\beta$-disk model might be a good nominee for the origin of planetary systems (e.g. AGS06).

\subsection{Role of polytropic exponent $\gamma$}
Another important input parameter of our model is $\gamma$ whose possible effects are examined in Figs. \ref{fig:alpha-gamma} and \ref{fig:beta-gamma}. We assume that $\alpha' = 0.1$, $\ell = 1.0$, $\beta'=10^{-3}$, $\dot{M}_{\rm infall}=2.0 \times 10^{-6} M_{\odot} / {\rm yr}$, $\Lambda_{0}=0.1$ and $\gamma=1.0$, $1.1$, $1.2$  with $l=1.0$ and $s=0.7$. As it is shown there, the radial velocity, surface density, vertical wind velocity, mass accretion rate and mass loss efficiency are significantly decreasing with $\gamma$ in both viscosity models. There is also a reduction to the rotational velocity in the presence of wind, and as $\gamma$ increases in $\alpha$-model or decreases in $\beta$-model, this reduction seems to be greater, under our restrictive boundary conditions. The disk aspect ratio $H/r$, angular momentum per unit mass $\jmath$, and also the Toomre instability parameter $Q$ (introduced in $\S \ref{subsec:q}$) are highly affected by the given parameter $\gamma$, and considerably increase with it. Thus, aside from the adopted viscosity prescription, the disks with larger $\gamma$s are thicker and gravitationally more stable.

\section{Summary and Outlook}

In the present study we have examined the influences of hydrodynamical winds from a geometrically thin disk rotating around the central object, taking account of the self-gravitational field of the disk gas with $\alpha$ and $\beta$-model for its viscosity.
We used the self-similar method to obtain the dimensionless form of the fluid equations, and then reduced them in the slow accretion limit.
In order to describe the evolution of our disk, we derived two sets of ODEs for two available models of viscosity.
We solved them numerically, by exploit of natural requirements and asymptotic
solutions near the origin and near the outer edge, as the boundary conditions.
Of course, it is important to keep in mind, we had the limitation
to select parameter $\gamma$ for essence of differential equations
and the fact that we seek just physical solutions (see MNU97 for constraints).

Some fraction of the accreted material and their angular momentum can be carried away by the wind. At the inner part of the disk, wind does not alter considerably the dynamical behaviors of the disk. However, at the outer part, where the wind appears more efficient, all physical variables would be much modified by wind. We presented the ratio of the total mass loss rate by wind to the mass accretion rate at each radius of the disk in all figures. There are some observations evidences of different systems which show that this ratio is around $0.1$ (e.g., K\"{o}nigl \& Pudritz 2000), which is in agreement with our results. Additionally, all the figures show consistency with our assumptions due to the slow accretion limit ($\S$3.1).

In spite of simplicity of our model in treating the wind and the disk itself, we think the presented semi-analytical results give us a better understanding of such a complicated system. Basically we had three main input parameters to control the physical properties of the wind in a phenomenological way, i.e. $s$, $l$ and $\Lambda_0$, and another input parameter related to the thermodynamics of the disk, i.e. $\gamma$. We did an extensive parameter study for a wide range of the input parameters and the main results are summarized as follows:

1. Radial dependence of the mass loss by wind was prescribed by a power law with exponent $s$. As this profile of mass loss becomes steeper with the radius, the accretion velocity is enhanced in particular at the outer layers of the disk. Additionally, the radial velocity increases with $\Lambda_0$. Naturally, in both viscosity models, if we keep all the input parameters fixed and decrease $s$ (because $0<x<1$) or increase $\Lambda_0$, then more mass is extracted from the disk by the wind. It means that more angular momentum is extracted by the wind in addition to the turbulent viscosity which implies a larger radial velocity as the solutions clearly show this behavior. Depending on the wind mechanism, value of $l$ is adopted in our model. Larger $l$ implies more efficient angular momentum extraction by the wind which leads to a more stable disk with larger radial velocity.

2. As the wind becomes stronger, the disk losses more mass and so, one should normally expect a reduction to the surface density of the disk. Consistent with this physical expectation, we showed that in the presence of wind, surface density decreases by decreasing $s$ or increasing $l$ or $\Lambda_0$ in both viscosity prescriptions.

3. In the model, the accretion rate depends on the radial velocity and the surface density. Although radial velocity increases, but surface density reduces in the presence of wind. As we analyzed the solutions of $\alpha$ and $\beta$ disks, the accretion rate reduces as the wind becomes stronger. Reduction to the accretion rate is not very sensitive to the value of $s$, but parameters $l$ and $\Lambda_0$ have a more significant effect to this reduction.

4. Since the structure of the disk is modified in the presence of wind according to the solutions, we also studied gravitational  stability of the disk via Toomre parameter. As the wind gets stronger, we see that Toomre parameters becomes larger which implies a more stable disk. However, dependence of Toomre parameter to the wind parameters are not at the same level. For example, Toomre parameter is not very sensitive to the exponent $s$. As it is shown in the figures, the gravitational instabilities in $\beta$-disks are more pronounced than $\alpha$-disks, even in the presence of wind. So, it might be anticipated that the $\beta$-model can better describe the planet formation around new-born stars. In the case of proto-planetary disks $\beta$-prescription yields the spectra that are considerably flatter than those due to non-self-gravitating disks, which is in a better agreement with observations (Abbassi \& Ghanbari 2008).

5. As may be inferred from the figures, $\beta$-disks typically have an aspect ratio $H/r$ smaller than $\alpha$-disks, and thus fall into the thin disk regime to a greater degree. However, this ratio increases as wind get stronger, irrespective of the viscosity prescription.

6. We also found that thermodynamics of the disk has a vital role even in our simplified picture in which a polytropic equation of state is used. With increasing $\gamma$, there are reductions to radial velocity, surface density, accretion rate and mass loss efficiency, but the angular momentum per unit mass increases. Moreover, the disk is geometrically thicker and gravitationally more stable as $\gamma$ increases.

The differences between $\alpha$ and $\beta$ models of viscosity prescription were predicated by Duschl et al. (2000), and indeed is confirmed by our results. In a global overview, as in AGS06, we have shown that in the outer part of the disk, where the self-gravity has an influential function, these models behave differently. They are somehow similarly affected by wind, though.
In the real accretion disks, there are several important
processes other than viscosity and wind, which are also expected to
transport angular momentum outward. It is also immediately clear that the changes in the boundary conditions affect the structure of the solutions. Many questions remain about the wind itself. For example, how is it driven and where does it leave the disk? We, however, did not consider the driving mechanisms of the wind. Given these facts, the treatments in the paper are very simplified, but sufficiently general to describe many of the disk-wind systems. All told, we believe that in order to obtain a better physical picture of such systems, more careful treatment is required and the analysis must be as deep as possible.
%
%
%
%

\section *{Acknowledgment}

The authors would like to thank the anonymous referee for the careful reading of the manuscript and his/her insightful and constructive comments. EN also wishes to thank SA and MS who supervised him on this project. This work has made extensive use of NASA's Astrophysical Data System Abstract Service (ADSABS).

\clearpage



\end{document}